\documentclass[runningheads]{llncs}
\usepackage{graphicx}
\begin{document}
\title{How the Far-Right Polarises Twitter:\\ 'Highjacking' Hashtags in Times of COVID-19}

\titlerunning{How the Far-Right Polarises Twitter}
%
\author{Philipp Darius\inst{1}  \and
Fabian Stephany\inst{2,3}\thanks{Corresponding Author: \email{fabian.stephany@oii.ox.ac.uk}}
}
\authorrunning{Darius and Stephany}

%
\institute{
Hertie School of Governance, Berlin \and
Oxford Internet Institute, University of Oxford, \and
Humboldt Institute for Internet and Society, Berlin
}

\maketitle              
\begin{abstract}
Twitter influences political debates. Phenomena like \textit{fake news} and \textit{hate speech} show that political discourse on micro-blogging can become strongly polarised by algorithmic enforcement of selective perception. Some political actors actively employ strategies to facilitate polarisation on Twitter, as past contributions show, via strategies of \textit{'hashjacking'}\footnote{The use of someone else’s hashtag in order to promote one's own social media agenda}. For the example of COVID-19 related hashtags and their retweet networks, we examine the case of partisan accounts of the German far-right party \textit{Alternative f\"ur Deutschland} (AfD) and their potential use of 'hashjacking' in May 2020. Our findings indicate that polarisation of political party hashtags has not changed significantly in the last two years. We see that right-wing partisans are actively and effectively polarising the discourse by 'hashjacking' COVID-19 related hashtags, like \#CoronaVirusDE or \#FlattenTheCurve. This polarisation strategy is dominated by the activity of a limited set of heavy users. The results underline the necessity to understand the dynamics of discourse polarisation, as an active political communication strategy of the far-right, by only a handful of very active accounts.\\

\end{abstract}
\section{Introduction}

The Coronavirus pandemic was accompanied by a so-called ‘infodemic’. This term refers to the acceleration of information and high uncertainty in the information environment, as for instance, in science, the media or policymaking. Moreover, there was a surge of dis- and misinformation and most debated ‘conspiracy theories/myths’ during the pandemic that additionally disrupted the media and information system. In this article, we show that far-right actors used long established strategies such as ‘hashjacking’ using political opponents hashtags and public hashtags to influence public opinion formation and leverage their own content. In this article, we assess polarisation strategies by partisans of the German right-wing populist party “Alternative für Deutschland” during two observation periods in 2018 and 2020. The analytical approach builds on community detection algorithms from social network analysis and logistic regression models to determine the likelihood of strategic hashtag hijacking or “hashjacking”. 

The results indicate that far-right users and politicians have been successful in establishing counterpublics on Twitter and were much more involved in linking their messages to wider discourses than partisans of other parties. By comparing the polarisation of party hashtags between 2018 and 2020, we identify political partisan groups and show that polarisation of party hashtags is relatively stable and highest for the far-right \#AFD. Moreover, our results indicate that Twitter debates on \#CoronaVirusDE and \#FlattenTheCurve are strongly polarised, too. The group of accounts spreading critical messages, questioning the existence of COVID-19 or the necessity of measures taken against the virus, is populated by partisans of the far-right. Within the far-right partisan movement around \#AFD a handful of heavy users are retweeting most of the content. It is important to know the dynamics of social media discourse, when engaging in it: A small group of  political partisans, in our case of the far-right AfD, systematically polarises discussions on public hashtags to influence public opinion formation and leverage their own content.

\section{Background}

Social platforms have become central hubs and fora for political debates, economic transactions, marketing and social communication. As a consequence, social platforms take an increasingly important role in the formation of public opinion, political campaigning and political news consumption. Yet, the media environment can be described as hybrid \cite{chadwick2017hybrid}, since television remains an important source of information and media consumption for large parts of the population \cite{newman2020digital}. Nevertheless, the share of citizens for which the Internet and social platforms represent the most important channels for political news consumption is increasing in many countries, e.g., Germany and the US.

For extreme parties and political opinions, social media offer additional channels for political communication in which extreme political actors do not need to follow the values and norms of traditional mass media and are thus able to spread their respective ideology \cite{engesser_populism_2017}. These alternative communication channels (social platforms and messengers) are particularly attractive for right-wing populist and radical-right politicians who often have a hostile attitude towards established media and less access to traditional media channels \cite{engesser_populism_2017}. Thus, right-wing populist actors and movements benefited disproportionately from the emergence of social media, since they could bypass traditional media gatekeepers and communicate directly with their respective target audiences \cite{stier_when_2017}. Now social platform companies increasingly take a role of being gatekeepers of information or “custodians of the Internet” \cite{gillespie2018custodians}. Besides, the direct contact to political actors and the represented ideologies enables the effect of a self-socialisation of citizens and users into right-wing populist beliefs and worldview \cite{kramer_populist_2017}. Consequently, social media also provide an opportunity for top-down claims of leadership for populist parties and politicians \cite{kramer_populist_2017}. 

As stated by the literature, social media have become an important political forum for the formation of political opinion and consequently also political persuasion and debate. When mapping these debates it is possible to identify different groups that arise due to ideologically aligned messaging behaviour \cite{conover_political_2011,conover_partisan_2012}. One strategy used by more or less organised political groups that is especially relevant for Twitter is the hijacking or ‘hashjacking’ of hashtags \cite{bode_candidate_2015,hadgu_political_2013,vandam_detecting_2016,darius2019hashjacking,knupfer_hijacking_2020}. This notion refers to the function of a hashtag to represent an issue, movement or organisation and the general expectation that the stream of information and messages pertaining to these hashtags resembles in some way the public opinion on that matter  \cite{bruns_use_2011,bruns_towards_2016,pond_riots_2019}. Consequently, the action of ‘hashjacking’ as using hashtags that were established by politically opposed groups or a general civil society discourse, is executed with the goal of quantitatively dominating the content referring to this specific hashtag \cite{darius2019hashjacking}. With regards to public hashtag debates, this strategy forms an adverse group or counter-public that is separated from the wider discourse and forms a separated cluster \cite{puschmann_information_nodate,xu_mapping_2020}.

\section{Hypotheses}

Revisiting our investigations on 'hashjacking' in 2018 \cite{darius2019hashjacking} and based on the presented literature on 'hashjacking' strategies, we formulate the following hypothesis.

As the share of citizens for which the Internet and social platforms represent the most important channels for political news consumption is increasing and political discourse is increasingly taking place on platforms like Twitter, we assume that debates around the major political parties in Germany remains to be polarised.

\begin{center}
    \textit{H1): The polarisation of retweet networks of political party hashtags did not change between 2018 and 2020.}
\end{center}

Likewise, we assume that civil society discourse on Twitter relating to issues like the
containment of the COVID-19 pandemic is subject to polarisation.

\begin{center}
    \textit{H2): Retweet networks of COVID-19 related hashtags show a significant degree of polarisation.}
\end{center}

As alternative communication channels (social platforms and messengers) are particularly attractive for right-wing populist and radical-right politicians, we assume that COVID-19 related hashtags, like \#CoronaVirusDE and \#FlattenTheCurve, have been systematically targeted by German far-right partisans of \#AFD.

\begin{center}
    \textit{H3): There is a significant activity of right-wing partisan accounts in contra-clusters of COVID-19 retweet networks.}
\end{center}

\section{Research Design}
The study is based on Twitter data that was collected by accessing Twitter's Rest API and using political party hashtags of German parties represented in the federal parliament (\#AFD, \#CDU, \#CSU, \#FDP, \#GRUENE, \#LINKE, \#SPD) and COVID-19 related hashtags (\#CoronaVirusDE and \#FlattenTheCurve) as a macro-level selection criterion. In total this study builds on a sample (n=101,765) of all public accounts using one or multiple of the selected political party hashtags between May 28th and June 4th in 2018 and 2020 on Twitter. The analysis focuses on a network approach and a visualisation of the retweet networks in Gephi using the Force2 layout algorithm \cite{jacomy2014forceatlas2} for each hashtag where retweeting creates a link (edge) between two accounts (nodes). Since the data was collected as separate streams of data pertaining to each hashtag, an account using several hashtags during the observation period in a retweet or being retweeted, will appear in each of the respective hashtag networks. 

Political discourses on Twitter show polarised or clustered structures due to the retweeting behaviour, as indicated by the literature. The analysis, consequently, focuses on the retweeting networks of the chosen hashtags. In a first step of analysis the modularity (community detection) algorithm \cite{blondel2008fast} assigns the nodes to different  communities based on the structural properties of the network graph. The cluster membership is indicated by the colour of nodes in Figure \ref{fig:network}. Thereafter, the interpretability of the clustering, as being in support of or opposition to a party, is controlled with a qualitative content analysis of the 50 most retweeted accounts similar to Neuendorf et al. \cite{neuendorf_content_2017}. This pro-/contra-polarisation of each party retweet network allows us to assign accounts as partisans of a specific party hashtag.

Based on previous invesitagations \cite{darius2019hashjacking}, we know that a high pro-party X \& contra-party Y association indicates 'hashjacking' strategies. Consequently, we use a logistic regression model to test all cluster combinations (as the likelihood to be in a contra-cluster of hashtag Y given a node was in the partisan cluster of party X). We decide to apply a logistic model for the assessment of cross-cluster heterophily, since our dependent variable is binary (location in contra-cluster) and the resulting odds are easy to interpret. Assuming there is no group that uses other party hashtags more frequently, users from all clusters should have the same odds to appear in the other network clusters. Thus, a high affiliation between two clusters in terms of their users being more likely to appear in both of them is an indication for strategic hashtag use or 'hashjacking'.

\section{Results}

\begin{figure}[h]
\begin{center}
\includegraphics[width=1\textwidth, keepaspectratio]{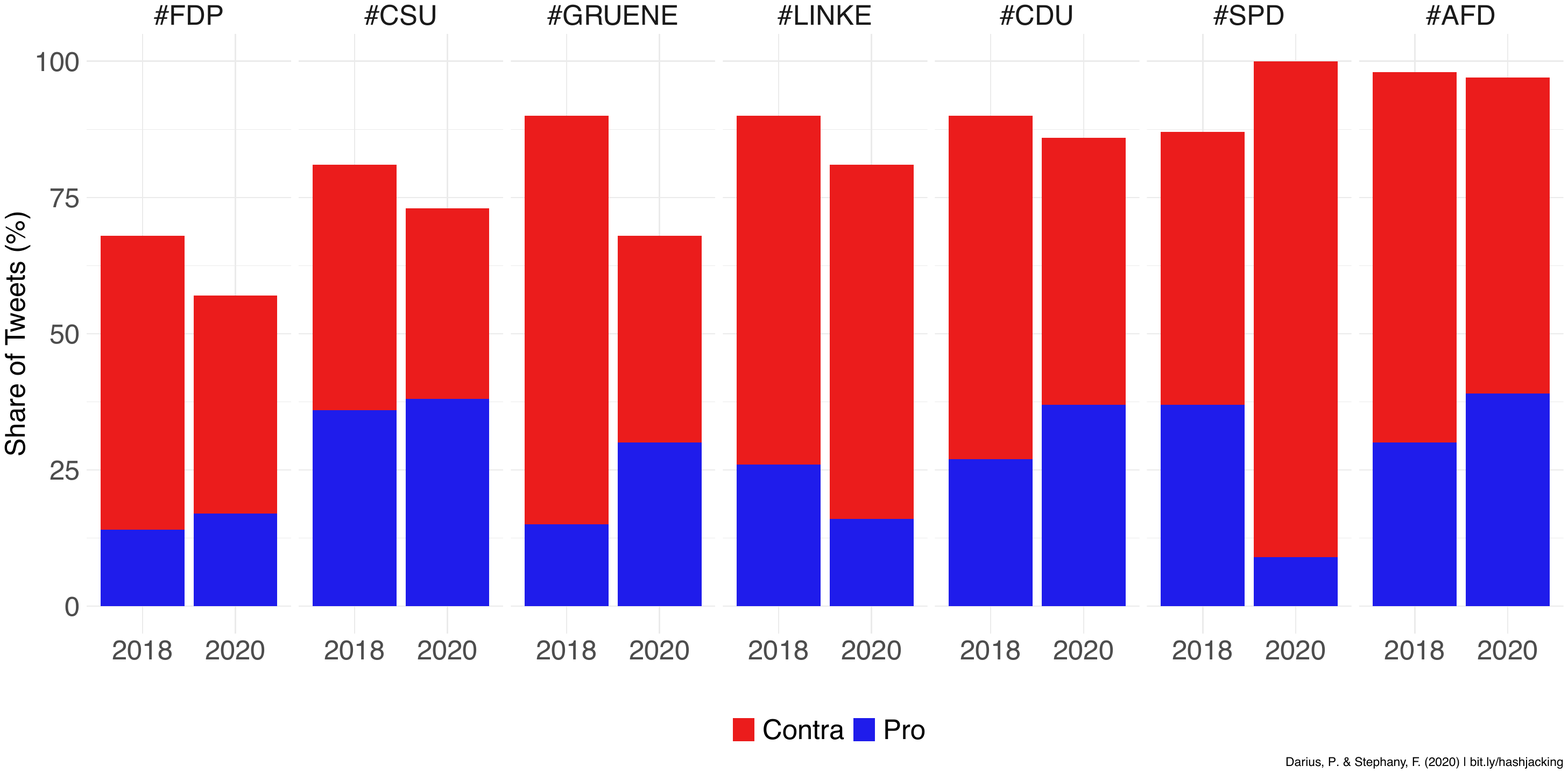}
\end{center}
\caption{\sf{In comparison to our data collected in May 2018, polarisation (the share of pro versus contra tweets) of the seven major German party hashtags has not changed significantly. It is still among highest for \#AFD, while the polarisation (growth of the contra-cluster) has likewise increased for the hashtag of the Socialdemocrats (\#SPD)}}
\label{fig:polarisation}
\end{figure}

In the first part of the analysis, we compare the polarisation of German political party hashtags, i.e., the share tweets in either a pro and contra-cluster, between our data sample collected 2018 and 2020. As illustrated in Figure \ref{fig:polarisation}, we observe that the share of polarised tweets did not change much between 2018 and 2020, thus confirming research hypothesis 1). Polarisation, in our definition, is still among highest for \#AFD, while the polarisation, i.e., growth of the contra-cluster, has likewise increased for the hashtag of the Socialdemocrats (\#SPD).

Secondly, similar to party hashtags, we construct retweeting networks for two COVID-19 related hashtags most popular in Germany in May 2020: \#CoronaVirusDE and \#FlattenTheCurve. With the example of \#CoronaVirusDE, Figure 2 illustrates the shape of these networks. After the application of a Louvain clustering \cite{de2011generalized} and modularity (community detection) algorithm, at least two different clusters emerge. Accounts coloured in dark blue in the bottom left of the network are centred around the account of \textit{@tagesschau}, Germany's major public-service national and international television news service. Similar to the accounts in the centre of the network (in light blue colours), users retweet news and governmental updates on COVID-19 in Germany. However, accounts the top-right cluster (in orange) have mainly shared tweets in which \#CoronaVirusDE had been used for questioning the existence of COVID-19 or the necessity of measures taken against the virus. Many of these accounts are partisans of \#AFD, as indicated in red. The strong clustering of the \#CoronaVirusDE retweet network (same holds for \#FlattenTheCurve) indicates a confirmation of hypothesis 2); retweet networks of COVID-19 related hashtags show a polarisation of users.

\begin{figure}[h]
\includegraphics[width=1.1\textwidth, keepaspectratio]{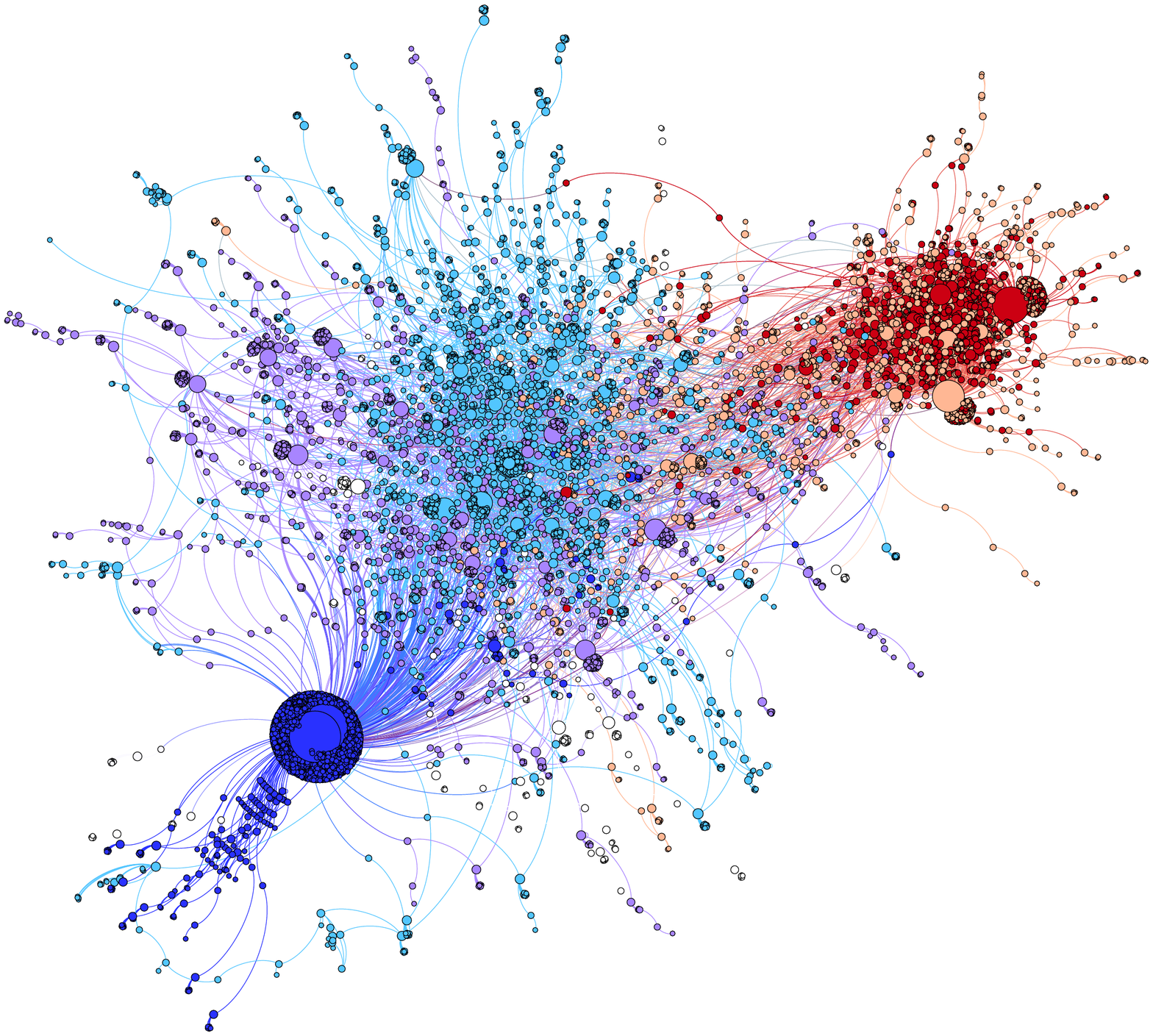}
\caption{\sf{In April 2020, users of the hashtag \#CoronaVirusDE are clearly clustered. Users top-right clusters (orange) have shared tweets questioning the existence of COVID-19 or the necessity of measures taken against the virus. 42 percent of these users and 69 of the 100 most active retweeters are partisans of the hashtag \#AFD (highlighted in red). We estimate the odds that a user retweeting \#CoronaVirusDE in critical tweets is a \#AFD partisan: $Odds(X\;  \epsilon\; Contra\; \#CoronaVirusDE\; |\; X\; \epsilon\; \#AFD\; partisan)$.}}
\label{fig:network}
\end{figure}

\begin{figure}
  \centering
  \textbf{(A)}
  \includegraphics[width=1\textwidth, keepaspectratio]{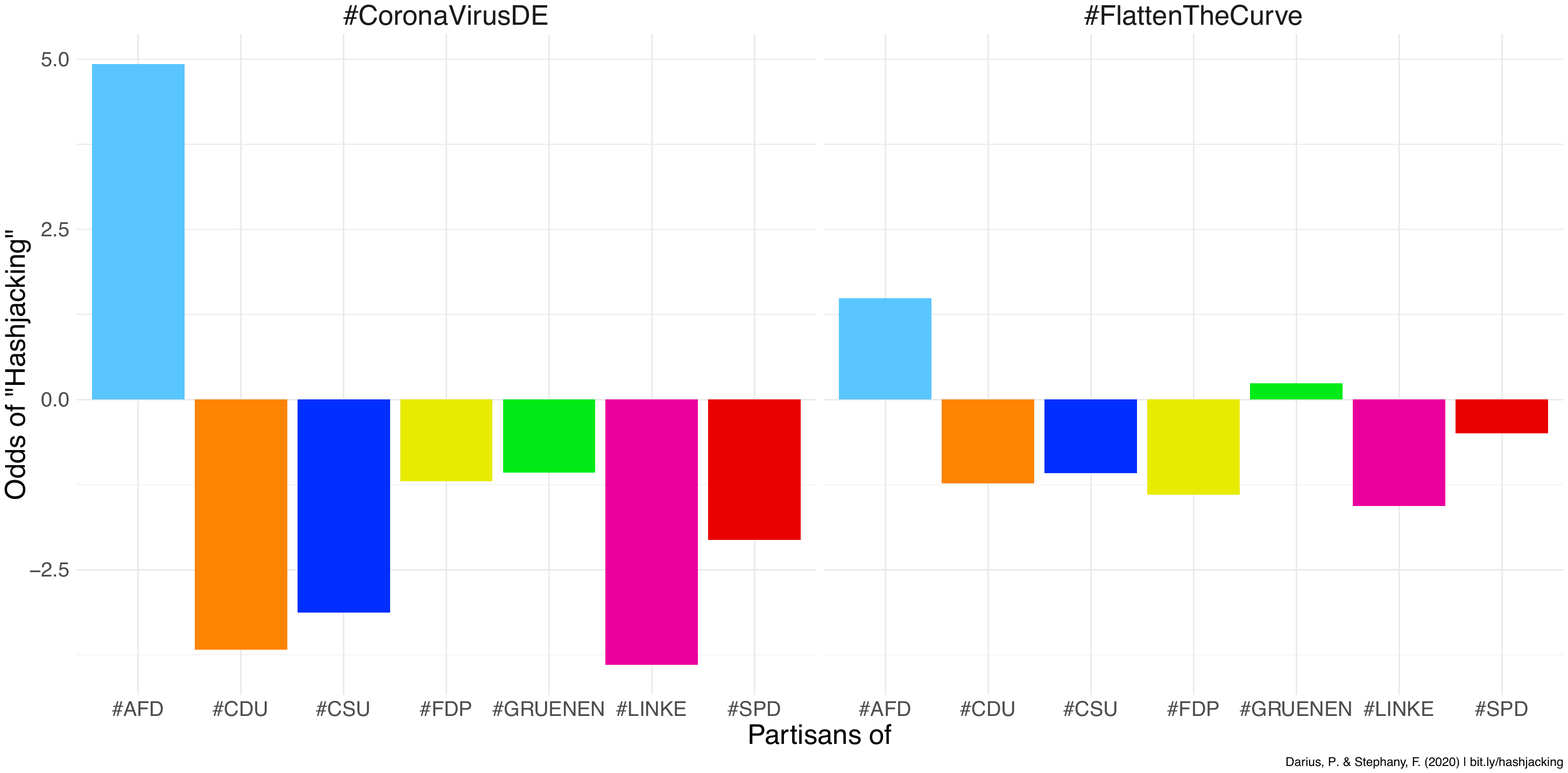}
  \textbf{(B)} 
  \includegraphics[width=1\textwidth, keepaspectratio]{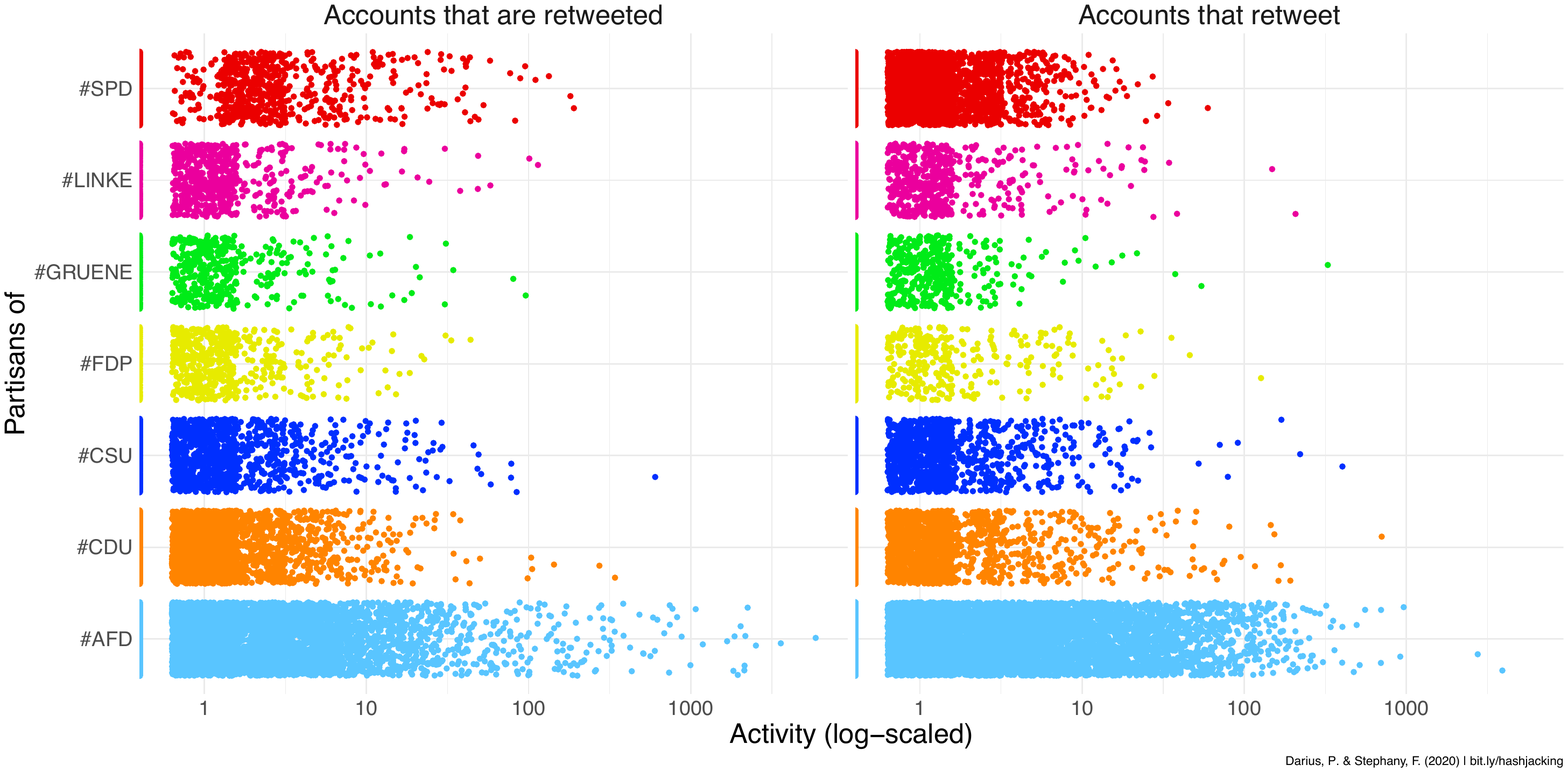}
  \caption{\sf{\textbf{(A)} For the retweet networks of \#CoronaVirusDE and \#FlattenTheCurve, partisans of \#AFD are five times, respectively 1.5 times more likely to appear in a contra-cluster than a user chosen at random. Partisans of other major Germany political party hashtags are less likely to occur in a contra-cluster. \textbf{(B)} The group of \#AFD partisans contains a set of heavy users. 39 percent of all \#AFD partisan retweets in our sample refer to one percent of \#AFD partisan accounts. Similarly, 50 percent of all retweets stem from the ten percent most active users.}}
  \label{fig:hashjacking}
\end{figure}

Based on the distribution of users in the retweet accounts of \#CoronaVirusDE and \#FlattenTheCurve we calculate conditional odds of 'hashjacking', i.e., the likeliness that an account in the contra-cluster of a given COVID-19 hashtag is part of a given partisan network, as shown in Figure \ref{fig:hashjacking}A). Here, we see that partisans of \#AFD are five times more likely to occur in the contra-cluster of \#CoronaVirusDE and 1.5 times more likely to be part of the contra-cluster of \#FlattenTheCurve than a user chosen at random. On the contrary, partisans of other party hashtags are less likely to appear in either of the contra-clusters. This finding delivers evidence confirming research hypothesis 3); a significant 'hashjacking' activity of right-wing partisans, which is higher than for partisan groups of other political party hashtags.

Lastly, we examine the activity distribution of individual partisan accounts. Figure \ref{fig:hashjacking}B) shows accounts across partisans by their activity of retweeting and being retweeted. For all partisan groups, we see a large share of users with very little activity (less than one retweet). However, the distribution of \#AFD partisans appears to be extremely long-tailed; it contains a set of heavy users. In fact, 39 percent of all \#AFD partisan retweets in our sample reference one percent of all \#AFD partisan accounts. Similarly, 50 percent of all retweets stem from ten percent of the most active users. The activity (retweeting) of far-right partisans is higher than for partisan groups of other political party hashtags.

Many of these heavy users contribute to the polarisation of COVID-19 related hashtags, like \#CoronaVirusDE. The \#AFD partisan accounts in the network in Figure \ref{fig:network} are highlighted in red. Their dominance in the sceptic contra-cluster (orange) is clearly visible. 42 percent of all users in the \#CoronaVirusDE contra-cluster are \#AFD partisans. In fact, 69 of the 100 most active users (in terms of retweets) in this cluster are partisans of \#AFD.


\section{Conclusion}
In this study we investigate the discourse polarisation on Twitter as a result of far-right partisans' application of a hashjacking strategy. For the case of COVID-19 related hashtags and German major political party hashtags, we find that partisans of the “Alternative f\"ur Deutschland” (\#AFD), systematically polarised discussions on \#CoronaVirusDE or \#FlattenTheCurve. Similarly, we find that polarisation on political party hashtags in Germany did not change much between 2018 and 2020. But for the exception of the Socialdemocrats' hashtag (\#SPD), with a growing opposition cluster, polarisation remains highest for the far-right. \#AFD partisans are much more likely to engage in retweeting on COVID-19 related hashtags questioning the existence of COVID-19 or the necessity of measures taken against the virus. At the same time, we highlight that the activity of far-right partisans is much driven by a handful of very active users. 50 percent of all retweets among \#AFD partisans stems from the ten percent most active accounts.
 
Our work validates the assumption of a high polarisation of debates related to COVID-19. Furthermore, we find evidence that this polarisation is strategically driven by partisans of the far-right. It takes only a small set of very active accounts to polarise retweeting debates on hashtags such as \#CoronaVirusDE or \#FlattenTheCurve. For users, it is important to be aware of these dynamics of social media discourse, when engaging in it. For platform providers, we show that the critical monitoring of a selected set of accounts can help to contain the spread of misinformation at large.


\newpage

%
%
%
\newpage
\bibliographystyle{ieeetr}
\bibliography{Bibliography}
\end{document}